\documentclass[12pt,headings=big,numbers=noenddot,DIV=14,a4paper]{scrartcl}%

\pdfoutput=1

\usepackage{amsmath,amssymb,amsfonts}
\usepackage[normal,font=small,labelfont=bf,labelsep=period]{caption}
\usepackage[pdftex]{color,graphicx}
\usepackage[english]{babel}
\usepackage[compress]{cite}
\addtokomafont{disposition}{\rmfamily\boldmath}

\usepackage[dvipsnames]{xcolor}
\definecolor{darkgreen}{rgb}{0,0.4,0}

\usepackage[linktoc=page,bookmarks=false,colorlinks=false,
linkbordercolor=RoyalBlue,citebordercolor=ForestGreen,urlbordercolor=CornflowerBlue]{hyperref}

\numberwithin{equation}{section}
\DeclareFontFamily{OT1}{pzc}{}
\DeclareFontShape{OT1}{pzc}{m}{it}{<-> s * [1.10] pzcmi7t}{}
\DeclareMathAlphabet{\mathpzc}{OT1}{pzc}{m}{it}

\title{\begin{flushright}
       \mbox{\small \rm SACLAY-T13/278}
       \end{flushright}
       \vskip 35pt
       \textcolor{black}{A bound on the charm chromo-EDM and its implications}}
\author{\normalsize Filippo Sala}
\date{\normalsize{\it Institut de Physique Th\'eorique, CNRS and CEA/Saclay, F-91191 Gif-sur-Yvette, France
\\ Scuola Normale Superiore and INFN, Piazza dei Cavalieri 7, 56126 Pisa, Italy\\
Theoretical Physics Group, Lawrence Berkeley National Laboratory, Berkeley, CA 94720
}}

\begin{document}
\begin{titlepage}

\maketitle
\thispagestyle{empty}

\begin{abstract}
\centerline{\bf Abstract}\medskip
\noindent
We derive bounds on the electric and chromo-electric dipole moments of the charm quark. The second one turns out to be particularly strong, and we quantify its impact on models that 
allow for a sizeable flavour violation in the up quark sector, like flavour alignment and Generic $U(2)^3$. In particular we show how the bounds coming from the charm and up CEDMs constrain the size of new physics contributions to direct flavour violation in $D$ decays. We also specialize our analysis to the cases of Supersymmetry with split families and composite Higgs models. The results exposed in this paper motivate both an increase in experimental sensitivity to fundamental hadronic dipoles, and a further exploration of the SM contribution to flavour violating $D$ decays.

 
\end{abstract}

\vfill
\noindent\line(1,0){188}\\\medskip
\footnotesize{E-mail: \href{mailto:filippo.sala@cea.fr}{filippo.sala@cea.fr} }

\end{titlepage}

\section{Introduction}
\label{sec1}

Electric dipole moments (EDMs) set stringent bounds on the CP structure of any new physics (NP) which becomes relevant at energies not far from the Fermi scale. An interesting question to ask is if and how one can exploit the current and foreseen experimental reach to constrain the flavour structure of such NP as well.
This issue becomes particularly relevant when the NP energy scale associated to the third generation is much lower than the one associated with the first two. This situation is typical of models which aim at evading collider and precision bounds while keeping the Fermi scale as natural as possible. In this class of theories, the new degrees of freedom related to the third generation often mediate the dominant contributions to the dipole moments of the light quarks. For quarks of the first generation, this immediately translates in a contribution to the EDMs of nucleons and nuclei. In this case, the non-observation of those EDMs sets bounds on flavour violating parameters relating the first and the third generation.
If also the second generation quarks were found to give relevant contributions to the EDMs of nucleons and/or nuclei, than one could also constrain flavour violation between the second and third generation. In this paper we show that this is actually possible, by computing the charm chromo-electric dipole moment (CEDM) contribution to the neutron EDM. We also show that the bound one derives in this way has interesting consequences for the flavour violating phenomenology of some models.

The current and foreseen experimental sensitivities to the electric dipole moments of the neutron, deuteron and mercury are summarized in Table \ref{tab:exp_bounds}. The quoted projection for $d_n$ is expected to be reached within a few years by more than one experiment, the one for $d_{\text{Hg}}$ by an upgrade of the same apparatus that sets the current bound. On the other hand, the experiment aiming at the measurement of $d_D$ is still in the proposal stage \footnote{For a more thorough discussion of future prospects see Section 7 of \cite{Hewett:2012ns} and references therein.}.
\begin{table}[h]
\renewcommand{\arraystretch}{1.3}
\centering
\begin{tabular}{r|ccc}
Observable & $d_n$ & $d_D$ & $d_{\text{Hg}}$ \\
\hline
Current bound & $2.9 \times 10^{-26}$  \cite{Baker:2006ts} & - & $3.1 \times 10^{-29}$ \cite{Griffith:2009zz} \\
Future sensitivity & $\sim 10^{-28}$ \cite{Bodek:2008gr,Altarev:2009zz,Baker:2010zza,Beck:2011gw,Altarev:2012uy} & $\sim 10^{-29}$  \cite{bnl:gov} & $\sim 10^{-30}$ \cite{Griffith:2009zz} 
\end{tabular}
\caption{Current bounds (90\% C.L. for $d_n$, 95\% C.L. for $d_{\text{Hg}}$) and expected sensitivities on the EDMs of the neutron, deuteron and mercury, in $e\,\text{cm}$.}
\label{tab:exp_bounds}
\end{table}

In the SM all the EDMs and CEDMs vanish exactly at the two-loop level\cite{Shabalin:1978rs}, the three-loop contributions have been evaluated in \cite{Khriplovich:1985jr,Czarnecki:1997bu} and, e.g. for the down quark, yield the estimate $d_d \simeq 10^{-34} e\, \text{cm}$. The neutron EDM is however dominated by long distance effects, the most recent estimation of them \cite{Mannel:2012qk} resulting in $d_n \simeq 10^{-31}  e\, \text{cm}$. This number is well below current and foreseen experimental sensitivities. Therefore $d_n$ remains a genuine probe of physics beyond the Standard Model.

This paper is organized as follows. In Section \ref{sec:bounds} we derive bounds on the electric and chromo-electric dipole moments of the charm quark. In Section \ref{sec:NP_implications} we discuss their implications for various NP models, both from an effective field theory (EFT) point of view (Sec. \ref{sec:EFT}) and in the specific cases of Supersymmetry (Sec. \ref{sec:SUSY}) and composite Higgs models (Sec. \ref{sec:CHM}). We summarize and conclude in Section \ref{sec:Summary}.


\section{Bounds on the charm quark dipole moments}
\label{sec:bounds}
In terms of fundamental dipoles, the electric dipole moments (EDMs) of the neutron\cite{Pospelov:2000bw},
deuteron \cite{Lebedev:2004va,deVries:2011re,deVries:2011an} and mercury \cite{Ellis:2008zy} read\footnote{A recent reevaluation of the neutron EDM \cite{Fuyuto:2012yf} sets a value which is smaller than the one used here, namely $ d_n = 0.79 d_d - 0.20 d_u + e (0.59 \tilde{d}_d + 0.30 \tilde{d}_u)$ (PQ-symmetric case, $w$ contribution ignored). The difference stems from having evaluated a parameter with the lattice instead of using QCD sum rules. For the mercury EDM, see also the recent error estimate of \cite{Jung:2013hka}, which makes the quark CEDMs impact compatible with zero.}:
\begin{align} 
 d_n = & (1 \pm 0.5) \big[1.4 (d_d - 0.25 d_u) + 1.1 e (\tilde{d}_d + 0.5 \tilde{d}_u) \big] \pm (22 \pm 10) \text{MeV}\,e\, w \,, \label{EDM_n}\\
 d_D = & - e (\tilde{d}_u - \tilde{d}_d) \big[4^{+7}_{-2}+(0.6 \pm0.3)\big] - (0.2 \pm 0.1) e (\tilde{d}_u + \tilde{d}_d) + \label{EDM_D} \\
 & + (0.5 \pm 0.3) (d_u + d_d) \pm e (22 \pm 10) \text{MeV}\, w \,, \nonumber \\
 d_{\text{Hg}} = & 7.2^{+14.4}_{-3.6} \times 10^{-3} \,e (\tilde{d}_u-\tilde{d}_d) + 10^{-2} d_e\,,
\end{align}
where $d_{u,d}, \tilde{d}_{u,d}$ are respectively the EDMs and CEDMs of the up and down quarks, $d_e$ is the electron EDM, and $w$ is the coefficient of the Weinberg operator. 
For $q = u,d,s,c,b,t$, they are defined via the following phenomenological Lagrangian
\begin{equation}
 \mathcal{L}_{\rm eff} = d_q \,\frac{1}{2} (\bar{q} \sigma_{\mu\nu} i \gamma_5 q) F^{\mu\nu} + \tilde{d}_q \,\frac{1}{2} (\bar{q} \sigma_{\mu\nu} T^a i \gamma_5 q) g_s G_a^{\mu\nu} + w \,\frac{1}{6} f^{abc} \epsilon^{\mu \nu \lambda \rho} G^a_{\mu\sigma} G^{b\,\sigma}_\nu G^c_{\lambda\rho}\,,
 \label{dipoles_def}
\end{equation}
where $\epsilon^{0123} = 1$.
The expressions \eqref{EDM_n} and \eqref{EDM_D} assume a PQ symmetry to get rid of the $\theta$ term. Ignoring this assumption would not only introduce a strong dependence on $\theta$, but also modify the one on the CEDMs. The CEDMs linear combination affecting the EDMs would change, but not the order of magnitude of their impact \cite{Pospelov:2000bw}. In studying the implications of the $d_n$ bound, in the rest of this paper we will conservatively use the values $0.5$ and $12$ MeV, respectively, for the coefficients $(1 \pm 0.5)$ and $(22 \pm 10)$ MeV in Eq. \eqref{EDM_n}.

The Weinberg operator in \eqref{dipoles_def} mixes via renormalization group (RG) evolution into the quarks EDMs and CEDMs, while the converse is not true. However, when in the running from high to low energies a quark $q$ is integrated out, its CEDM gives the following threshold correction to the Weinberg operator at one-loop level \cite{Chang:1990jv,Boyd:1990bx,Dine:1990pf}
\begin{equation}
w = \frac{g_s^3}{32 \pi^2} \frac{\tilde{d}_q}{m_q}\,,
 \label{Weinberg_threshold}
\end{equation}
where all the parameters are evaluated at the mass of the quark. The uncertainty from going to higher loops in \eqref{Weinberg_threshold} can be estimated to be at the level of $8 \, \alpha_s(m_q)/4 \pi$, about $25 \%$ for $q=c$, where 8 is a colour factor. The subsequent running makes also the lighter quarks dipole moments sensitive to $\tilde{d}_q$. In terms of the charm CEDM evaluated at the scale $m_c$, $w$ and the dipoles $d_{u,d}$, $\tilde{d}_{u,d}$ at the hadronic scale of 1 GeV read
\begin{equation}\begin{array}{ll}
 \tilde{d}_u = 1.7 \times 10^{-6} \,\tilde{d}_c\,, \qquad \quad & d_u = -5.9 \times 10^{-8} e \,\tilde{d}_c \,,\\
 \vspace{-.2 cm} & \\
 \tilde{d}_d = 3.5 \times 10^{-6} \,\tilde{d}_c\,, \qquad \quad & d_d = 6.2 \times 10^{-8} e \,\tilde{d}_c \,,\\
 \vspace{-.2 cm} & \\
 w = 2.3 \times 10^{-2} \text{GeV}^{-1} \,\tilde{d}_c\,.& 
 \label{charm_CEDM_into}
\end{array}\end{equation}
In deriving \eqref{charm_CEDM_into} we have used the running from \cite{Braaten:1990gq,Degrassi:2005zd} at one-loop. The relevant running of the Weinberg operator at two-loops is, to our knowledge, unknown. Moreover, in the extraction of a bound for $\tilde{d}_c$, the impact of the up and down EDMs is subleading with respect to the one of $w$. This is evident by inserting \eqref{charm_CEDM_into} into \eqref{EDM_n} and \eqref{EDM_D}, and makes the known two-loop running unnecessary.

The experimental bound on $d_n$ then implies
\begin{equation}
|\tilde{d}_c| \lesssim 1.0 \times 10^{-22} \text{cm}\,,
\label{eq:CEDMc_bound} 
\end{equation}
or, equivalently, $m_c |\tilde{d}_c| \lesssim 6.7 \times 10^{-9}$. This is to be compared to the previous and only bound existing in the literature, $|\tilde{d}_c| \lesssim 3 \times 10^{-14}$ cm, obtained from $\psi^\prime \to \psi \pi^+ \pi^-$ at the Beijing spectrometer \cite{Kuang:2012wp}.
As already said, the bound \eqref{eq:CEDMc_bound} comes mainly from the direct contribution of $w$ to $d_n$.
The mercury EDM bound thus yield a much weaker constraint on $\tilde{d}_c$, than the one set by $d_n$.
An analysis analogous to the one we performed here can be carried out also for the bottom and top CEDMs, as was done in \cite{Chang:1990jv} and \cite{Kamenik:2011dk}.
As a cross-check of our derivation, we verified that our procedure reproduces their results.

The indirect constraints on the charm EDM are weaker. They can be derived from both the mixing of $d_c$ into $d_d$ via electroweak running, and from the $d_c$ contribution to $B\to X_s \gamma$. In the first case, using \cite{CorderoCid:2007uc} for the running and the bound on $d_n$, one gets

\begin{equation}
|d_c| \lesssim 4.4 \times 10^{-17} e\, \text{cm} \,,
\label{eq:EDMc_bound_dn} 
\end{equation}
where again $d_c$ is evaluated at the charm mass scale.
In the case of $B\to X_s \gamma$, the contribution of $d_c$ is relevant since it has the same loop and CKM suppressions of the Standard Model one ($|V_{cb}| \simeq |V_{ts}|$). To derive it, we use \cite{Hewett:1993em} for the charm dipole contribution to the Wilson coefficient $C_{7\gamma}$, and \cite{Buras:2011zb} for the dependence of BR$(B\to X_s \gamma)$ on $C_{7\gamma}$. In explicit models one generically expects a charm magnetic dipole moment, of size similar to $d_c$, to be generated. However the sensitivity of $C_7$ to it is more than one order of magnitude smaller than the one to $d_c$, and we ignore it here for simplicity, like we do with other possible NP contributions. We then obtain
\begin{equation}
|d_c| \lesssim 3.4 \times 10^{-16} e\, \text{cm} \,,
\label{eq:EDMc_bound_Bsgamma} 
\end{equation}
where we have used the experimental world average \cite{Amhis:2012bh} BR$(B\to X_s\gamma) = (3.43 \pm 0.22) \times 10^{-4}$, and imposed the bound at 2$\sigma$ (where the uncertainty of the SM contribution BR$(B\to X_s\gamma)_{\rm SM} = (3.15\pm0.23) \times 10^{-4}$ \cite{Misiak:2006zs} has been added in quadrature). 


\section{Implications for New Physics}
\label{sec:NP_implications}

It can be convenient to express NP contributions to the EDM and CEDM of a given quark $q$ in terms of the following high scale effective Lagrangian
\begin{equation}\label{EDM}
\mathcal{L}_\text{dip} = \frac{m_t}{\Lambda^2} \,\xi_q \,\left[
 c_q  (\bar q_L\sigma_{\mu\nu} q_R) eF^{\mu\nu} +
\tilde c_q (\bar q_L\sigma_{\mu\nu}T^a q_R)  g_s G^{\mu\nu}_a \right]
+\text{h.c.} \,,
\end{equation}
where $c_q, \tilde c_q$ are coefficents of order one, and $\xi_q$ are suppression factors, all depending on the specific model and in principle complex. With these definitions, the quark EDMs and CEDMs read
\begin{equation}
d_{q} = 2 e \, \frac{m_t}{\Lambda^2}\, \text{Im}(c_q \xi_q),\qquad \tilde d_q = 2 \, \frac{m_t}{\Lambda^2}\, \text{Im}(\tilde c_q \xi_q)\,,
\end{equation}

\subsection{Size of the bounds in EFT}
Imposing $d_n < 2.9 \times 10^{-26}$ $e$ cm and considering one operator at a time in \eqref{EDM}, for $\Lambda = 1$ TeV we find
\begin{align}
 \text{Im}(\tilde c_u \xi_u) &\lesssim 1.3 \times 10^{-8}, & \text{Im}(c_u \xi_u) \lesssim 3.3 \times 10^{-8},\label{EDMu_bounds}\\
 \text{Im}(\tilde c_d \xi_d) &\lesssim 8.4 \times 10^{-9}, & \text{Im}(c_d \xi_d) \lesssim 6.5 \times 10^{-9},\\
 \text{Im}(\tilde c_c \xi_c) &\lesssim 1.8 \times 10^{-5}, & \label{EDMc_bounds}\\
 \text{Im}(\tilde c_b \xi_b) &\lesssim 1.7 \times 10^{-4}, & \label{EDMb_bounds}\\
 \text{Im}(\tilde c_t \xi_t) &\lesssim 3.3 \times 10^{-2}, & \label{EDMt_bounds}
\end{align}
where all the coefficients are evaluated at the scale $\Lambda = 1$ TeV. 

Notice that the 4 fermion operator contributions to the EDMs \cite{Hisano:2012cc} have been ignored. Given the uncertainties present in casting the bounds, this approximation is justified in those models where such operators are not enhanced with respect to the dipole ones.
This happens for example in Supersymmetry, where they arise at loop level, or in composite Higgs models with partial compositeness, where they appear at tree level but their coefficients are further suppressed, with respect to the dipole operators ones, by an extra light quark Yukawa coupling. 

\subsection{Interplay with bounds from flavour violating processes}
\label{sec:EFT}

The new bound we derived can be relevant for models allowing for a sizeable flavour violation in the right-handed up quark sector, while at the same time providing a large splitting between the energy scales associated with the third and the first two generations of quarks. Such a scenario is favoured by naturalness arguments when combined with current direct NP searches, and consistent with data due to the fact that the stronger constraints in flavour violation come from processes involving down quarks. 
Explicit realizations are models of flavour alignment (see e.g. \cite{Nir:1993mx}), composite Higgs models (CHM) with an anarchic flavour structure, or Generic $U(2)^3$ \cite{Barbieri:2012bh}.

In such models, measurements of CP asymmetries in processes like $D \to \pi \pi$ and $D \to K K$ are among the most stringent probes of flavour violation in the up quark sector.
This is true in particular for chromo-magnetic dipole operators of both chiralities, that are instead less efficiently constrained by $D-\bar D$ mixing or $\epsilon^{\prime}_K$ \cite{Isidori:2011qw}.
We write the high scale effective Lagrangian contributing to such processes as
\begin{equation}
\mathcal{L}^{\Delta C=1}_\text{mag} = \frac{m_t}{\Lambda^2} \left[ c_D \xi_8 \mathcal{O}_8
+
c_D^\prime \xi_8^\prime \mathcal{O}_8'
\right] + \text{h.c.}\\
\label{eq:Leff_Ddecay}
\end{equation}
where
\begin{equation}
\mathcal{O}_8 = (\bar u_L\sigma_{\mu\nu}T^a c_R)g_s G^{\mu\nu}_a,\qquad \mathcal{O}_8' = (\bar u_R\sigma_{\mu\nu}T^a c_L)g_s G^{\mu\nu}_a,
\end{equation}
$c_D, c_D^{\prime}$ are coefficients of order one, and $\xi_8, \xi_8^\prime$ are suppression factors, all depending on the model and in principle complex.
The most recent measurement of the CP asymmetry in $D$ decays is\cite{Aaij:2013bra} $\Delta\rm A_{\rm CP} = \rm A_{\rm CP}(K^+K^-) - \rm A_{\rm CP}(\pi^+\pi^-) = \big( 4.9 \pm 3~(\rm stat) \pm 1.4 ~(\rm syst) \big) \times 10^{-3}$, yielding to the world average \cite{Amhis:2012bh} $\Delta\rm A_{\rm CP} = (-3.29 \pm 1.21) \times 10^{-3}$. The Standard Model contribution could possibly account for such a value, however its determination is still object of intensive discussion, see e.g.\cite{Pirtskhalava:2011va,Cheng:2012wr,Brod:2012ud,Isidori:2012yx}. Our approach is therefore to require the NP contribution to be smaller than the average central value: following the analysis of \cite{Isidori:2011qw} and considering one operator at a time, we find that this implies, for $\Lambda = 1$ TeV,
\begin{equation}
 \label{eq:DAcp_bound}
 \text{Im}(c_D^{(\prime)} \xi_8^{(\prime)}) < 3.8 \times 10^{-6}\, .
\end{equation}
It is important to keep in mind that the above bound is plagued by $O(1)$ uncertainties due to the poor knowledge of the matrix elements of $\mathcal{O}_8$ and $\mathcal{O}_8'$.

As stated in the introduction, we are interested in models where the degrees of freedom associated with the third generation are those giving the dominant contribution to the operators in \eqref{EDM} and \eqref{eq:Leff_Ddecay}. This translates in the assumptions
\begin{equation}
 \xi_8 = W^L_{u3} W^R_{3c}, \qquad\xi_8^{\prime} = W^L_{c3} W^R_{3u}\,,
 \label{eq:xi_DeltaAcp}
\end{equation}
and, for the dipole moments
\begin{equation}
 \xi_q = W^L_{q3} W^R_{3q}\,,
 \label{eq:xi_EDMs}
\end{equation}
where $W^L_{i3}$ and $W^R_{3i}$ are flavour violating parameters that quantify the communication between the i$^{\rm th}$ generation of quarks, and the new degrees of freedom associated with the third generation. For instance in Supersymmetry, if gluino contributions dominate, they are the matrices in flavour space in the gluino-quark-squark vertices.
Notice that everywhere $\Lambda$ is the energy scale associated with the third generation quarks, and that the phases of the parameters in \eqref{eq:xi_DeltaAcp}, \eqref{eq:xi_EDMs} are flavour violating ones.
\begin{itemize}
 \item The first important observation is that $\xi_u \xi_c = \xi_8 \xi_8^{\prime}$. In the absence of a direct constraint on $\xi_c$, it was the bound from $\Delta$A$_{\rm CP}$ that allowed to set the stronger constraint on that combination of parameters. Now, as one can see by taking the product of \eqref{EDMu_bounds} and \eqref{EDMc_bounds}, and of \eqref{eq:DAcp_bound} with itself, the EDM of the neutron is already setting the stronger bound by a factor of $\sim 60$. This conclusion will be strengthened by the foreseen experimental sensitivities, in the absence of improvements in the understanding of the SM contribution to $\Delta$A$_{\text{CP}}$.
 \item The above generic situation can be specialized to the case of $W^L_{q3} \simeq V_{qb}$, with $V$ the CKM matrix, as typical of models of alignment
 . We now assume, for simplicity, maximal phases and all the $O(1)$ coefficients to be one. In this case, the bounds from $\Delta$A$_{\text{CP}}$ imply
 \begin{equation}
  |W^R_{3c}| < 1.1 \times 10^{-3}, \qquad |W^R_{3u}| < 9.2 \times 10^{-5}\,,
 \end{equation}
 and those from the charm and up CEDMs require, respectively,
 \begin{equation}
  |W^R_{3c}| < 4.4 \times 10^{-4}, \qquad |W^R_{3u}| < 3.7 \times 10^{-6}\,.
 \end{equation}
 where again we have chosen a NP scale $\Lambda = 1$ TeV, and considered one operator at a time.
 Without considering the contributions from the charm CEDM computed in this paper, one could have saturated the $\Delta$A$_{\text{CP}}$ measured value without being in conflict with the EDMs constraints, via requiring a very small $W^R_{3u}$, see e.g. \cite{Giudice:2012qq}. Now this possibility is challenged and, with the forseen experimental sensitivities, in these models the neutron EDM will become by far the most powerful observable to probe the flavour violating parameters in \eqref{eq:xi_DeltaAcp} and \eqref{eq:xi_EDMs}. This conclusion would be strengthened by more than an order of magnitude (totalizing a $\sim 10^3$ better sensitivity to $|W^R_{3c}|$ with respect to $\Delta$A$_{\text{CP}}$) if the deuteron EDM will be measured with a precision of $\sim 10^{-29} e$~cm.
 We stress that all these bounds should be considered as $O(1)$ limits, barring finetunings of the unknown coefficients and  overall phases in front of the operators considered here. This implies, for example, that formally there is the possibility to make the phases entering the CEDMs small so to be in agreement with the bounds \eqref{EDMu_bounds} and \eqref{EDMc_bounds}, while keeping larger the ones relevant to $\Delta$A$_{\text{CP}}$ and invalidate the above conclusion.
 \item In Generic $U(2)^3$ models, one has
 \begin{equation}
  W^L_{q3} = V_{qb}, \quad W^R_{c3} = V_{cb} \epsilon_c, \quad W^R_{u3} = V_{ub} \epsilon_u\,,
 \end{equation}
 where $\epsilon_u < \epsilon_c < 1$ are suppression parameters related to the breaking of $U(2)^3$ symmetry in the right quark sector.\footnote{In terms of the notation of Ref. \cite{Barbieri:2012bh}, $\epsilon_u = \dfrac{s_R^u}{s_L^u} \dfrac{\epsilon_R^u}{\epsilon_L}$ and $\epsilon_c = \dfrac{\epsilon_R^u}{\epsilon_L}$.}
 In the case of maximal phases and $O(1)$ coefficients equal to one, again considering one operator at a time, the bounds from $\Delta$A$_{\text{CP}}$ now imply
 \begin{equation}
  \epsilon_c < 2.6 \times 10^{-2}, \qquad \epsilon_u < 2.6 \times 10^{-2}\,,
 \end{equation}
 and those for the charm and up CEDMs require, respectively,
 \begin{equation}
  \epsilon_c < 1.1 \times 10^{-2},\qquad \epsilon_u < 1.0 \times 10^{-3}\,.
 \end{equation}
 Like before, the EDMs are starting to become more sensitive to the parameter $\epsilon_c$ than direct CP violation in charm decays, and will become the best observable to probe the amount of $U(2)^3$ breaking in the up-right quark sector. In this scenario, the flavour symmetry imposes the following relations among the $O(1)$ complex coefficients: $c_D = \tilde{c}_c$ and $c_D^\prime =  \tilde{c}_u$ (see Appendix A.2 of \cite{Barbieri:2012bh}). Thus, contrary to the previous case, in Generic $U(2)^3$ it is not possible to play with the order one parameters and phases to avoid the above conclusions. 
\end{itemize}
A remark is in order to avoid possible confusion. In \cite{Mannel:2012hb} direct CP violation in $D$ meson decays was related to the neutron EDM. The result was that the same $\Delta C = 1$ operators inducing $\Delta$A$_{\rm CP}$ at a level compatible with the measured value, also induce a contribution to $d_n$. This contribution is obtained by long distance effects at tree-level, in analogy with the dominant SM contribution by the same authors \cite{Mannel:2012qk}, and its size is at most one order of magnitude below the current experimental sensitivity (and now even smaller, in light of the new $\Delta$A$_{\rm CP}$ measurement).
Here we pursue a different analysis, namely we identify a class of models where a sizeable contribution to $\Delta$A$_{\rm CP}$ is accompained by flavour conserving CP-violating operators, and study the impact of the last ones on $d_n$, that was not considered in \cite{Mannel:2012hb}. The contribution to $d_n$ that we find, in these explicit models, is more than an order of magnitude larger than the model independent one obtained in \cite{Mannel:2012hb}.


\subsection{Supersymmetry with split families}
\label{sec:SUSY}
Split-families SUSY (often referred to as ``Natural SUSY'') \cite{Dimopoulos:1995mi,Cohen:1996vb,Barbieri:2009ev,Papucci:2011wy} is an explicit realization of the situation described in the previous section.
The dominant contributions to the Wilson coefficients defined in \eqref{EDM} and \eqref{eq:Leff_Ddecay} are induced by gluino-stop loops, and read
\begin{equation}
 \frac{c_D}{\Lambda^2} = \frac{c_D^\prime}{\Lambda^2} =\frac{\tilde c_c}{\Lambda^2} = \frac{\tilde c_u}{\Lambda^2}= \frac{\alpha_s}{4 \pi} \frac{1}{m_{\tilde g}^2} \frac{A_t - \mu \cot \beta}{m_{\tilde t}} \frac{5}{36}\, g_8(x_{gt})\,,
 \label{SUSY_C8}
\end{equation}
where $x_{gt} = \dfrac{m_{\tilde g}^2}{m_{\tilde t}^2}$ and
\begin{equation}
 g_8(x) = x^{\frac{3}{2}} \left(\frac{12}{5} \,\frac{11+x}{(x-1)^3} -\frac{6}{5}\,\frac{9+16 x-x^2}{(x-1)^4}\log{x}\right) \,,\qquad g_8(1) = 1\,.
\end{equation}
In the suppression factors $\xi_8$, $\xi_8^\prime$ and $\xi_q$, the elements $W^{L(R)}_{q3}$ are those of the mixing matrices entering the gluino-quark-squark vertices of the respective chirality, which are responsible for the flavour violation.
Fixing for illustrative purposes $m_{\tilde{g}} = 2 m_{\tilde{t}}$ and assuming maximal phases, the bounds from the CEDMs of the up and charm quarks read, respectively
\begin{equation}\begin{array}{ll}
\left| W_{tu}^R \dfrac{W_{tu}^L}{V_{ub}} \left( \dfrac{A_t - \mu/\tan\beta}{m_{\tilde{t}}}\right)\right| \left(\dfrac{1.5 \text{TeV}}{m_{\tilde{g}}} \right)^2 &\lesssim 6.5 \times 10^{-3}\,, \\
\vspace{-0.2 cm}& \\
\left| W_{tc}^R \dfrac{W_{tc}^L}{V_{cb}} \left( \dfrac{A_t - \mu/\tan\beta}{m_{\tilde{t}}}\right)\right| \left(\dfrac{1.5 \text{TeV}}{m_{\tilde{g}}} \right)^2 &\lesssim 0.77\,,
 \label{eq:SUSY_CEDMS_bounds}
\end{array} \end{equation}
to be compared with the ones coming from $\Delta$A$_{CP}$
\begin{equation}\begin{array}{ll}
\left| W_{tu}^R \dfrac{W_{tc}^L}{V_{cb}} \left( \dfrac{A_t - \mu/\tan\beta}{m_{\tilde{t}}}\right)\right| \left(\dfrac{1.5 \text{TeV}}{m_{\tilde{g}}} \right)^2 &\lesssim 0.16\,, \\
\vspace{-0.2 cm}& \\
\left| W_{tc}^R \dfrac{W_{tu}^L}{V_{ub}} \left( \dfrac{A_t - \mu/\tan\beta}{m_{\tilde{t}}}\right)\right| \left(\dfrac{1.5 \text{TeV}}{m_{\tilde{g}}} \right)^2 &\lesssim 1.9\,.
 \label{eq:SUSY_DeltaAcp_bounds}
\end{array} \end{equation}
Choosing instead $m_{\tilde{t}} = m_{\tilde{g}}$ one would obtain bounds weaker by a factor of $\sim 1.3$.

\begin{figure}
\begin{center}
  \includegraphics[width=9 cm]{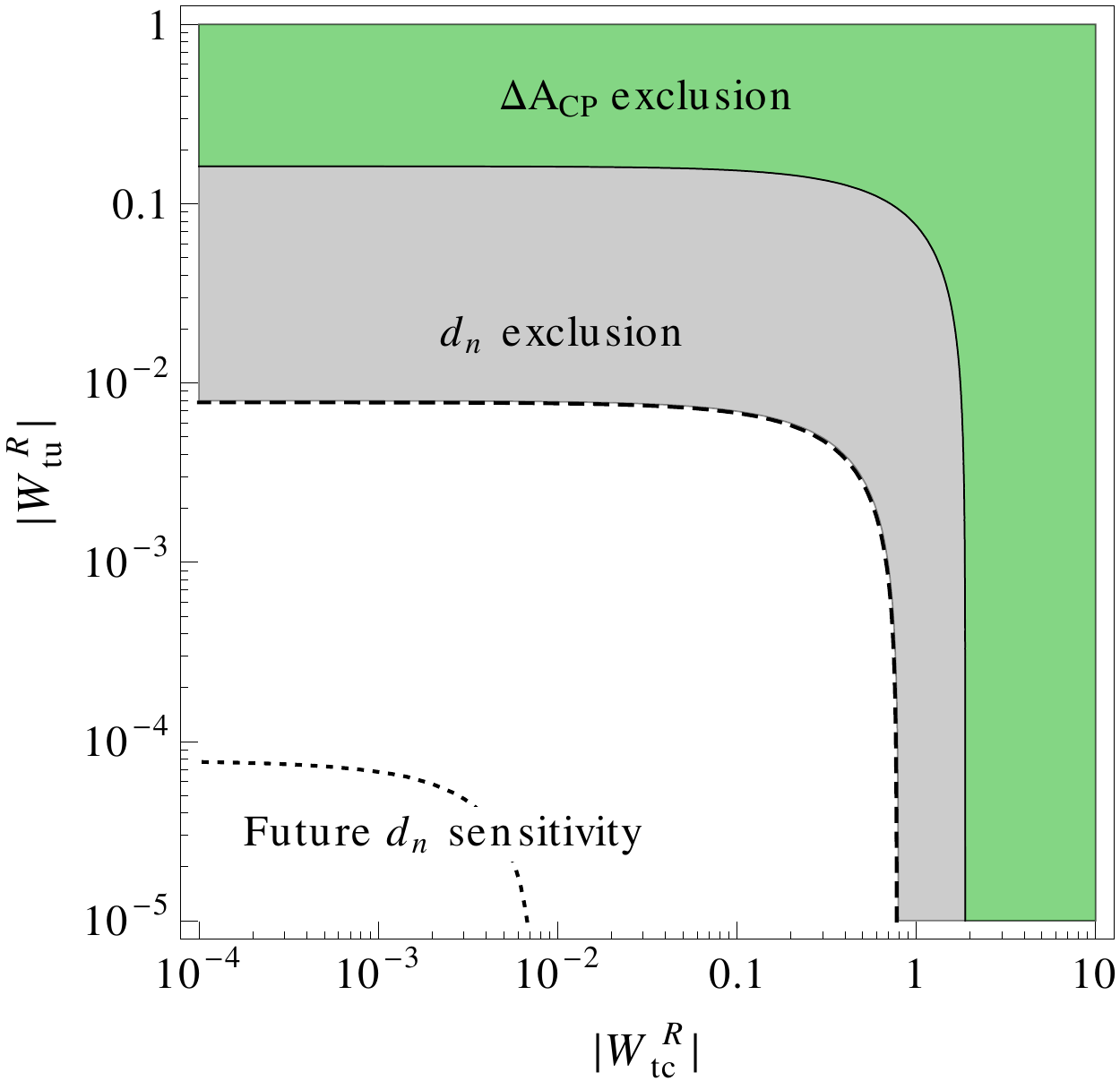}
\end{center}
  \caption{The lines represent the experimental sensitivities to the flavour violating matrix elements in split-families SUSY, the shaded region are currently excluded. Dashed: current neutron EDM. Continuous: Direct CP asymmetry in $D$ decays. Dotted: projected neutron EDM.}
\label{fig:SUSY_bounds}
 \end{figure}

In split-families SUSY one can improve the robustness of the previous bounds by taking into account all the dominant contributions to $d_n$. Under some assumptions that will be discussed, it is in fact sufficient to add to the previous picture the up electric dipole moment $d_u$.
To see this, let us first consider the bounds on the top and bottom CEDMs, \eqref{EDMt_bounds} and \eqref{EDMb_bounds}. Again the Supersymmetric contribution to them is dominated by gluino-squark loops, and it reads as the one in Eq. \eqref{SUSY_C8}, with the appropriate squark mass and mixing substitution for the bottom case. In addition, the suppression factor for the top case reads $W_{tt}^R W_{tt}^L$, the one for the bottom $y_b/y_t W_{bb}^R W_{bb}^L$. The bounds \eqref{EDMb_bounds} and \eqref{EDMt_bounds} then imply
\begin{equation}\begin{array}{ll}
\left|W_{bb}^R W_{bb}^L \left( \dfrac{A_b - \mu \tan\beta}{m_{\tilde{t}}}\right)\right| \left(\dfrac{1.5 \text{TeV}}{m_{\tilde{g}}} \right)^2 & < 18\,, \\
\vspace{-0.2 cm}& \\
\left|W_{tt}^R W_{tt}^L \left( \dfrac{A_t - \mu/\tan\beta}{m_{\tilde{t}}}\right)\right| \left(\dfrac{1.5 \text{TeV}}{m_{\tilde{g}}} \right)^2 & < 59\,.
 \label{eq:SUSYbound_bottom_top}
\end{array} \end{equation}
Thus it is safe to neglect the top and bottom CEDMs contribution to $d_n$ for values of the matrix elements of order one.
Let us now come to the contribution from the down quark EDM and CEDM. First notice that, with respect to the up quark (C)EDMs, they are suppressed by a bottom yukawa coupling, being proportional to $y_b/y_t W_{bd}^L W_{bd}^R$. Also, the $\epsilon_k$ parameter constrains the size of the combination $W_{bs}^L W_{bs}^R W_{bd}^L W_{bd}^R$ to be much smaller than the corresponding one in the up sector, if sbottom and stops have similar masses. In light of these observations, we assume a negligible down quark contribution to the neutron EDM. One is left then with $d_u$, $\tilde{d}_u$ and $\tilde{d}_c$ as dominant contributions to $d_n$.
The coefficient of the up-quark electric dipole moment, in the notation of \eqref{EDM}, reads
\begin{equation}
 \frac{c_u}{\Lambda^2} = \frac{\alpha_s}{4 \pi} \frac{1}{m_{\tilde g}^2} \frac{A_t - \mu \cot \beta}{m_{\tilde t}} \frac{1}{27}\, g_7(x_{gt})\,,
\end{equation}
where
\begin{equation}
 g_7(x) = x^{\frac{3}{2}} \left(-6 \,\frac{1+5 x}{(x-1)^3} +12\,\frac{x (2+x)}{(x-1)^4}\log{x}\right) \,,\qquad g_7(1) = 1\,.
\end{equation}
The neutron EDM can then be written in the compact form

\begin{equation}
 \frac{d_n}{2.9 \cdot 10^{-26} e \text{cm}} = \left(130 \cdot |W_{tu}^R| \left|\dfrac{W_{tu}^L}{V_{ub}}\right| s_u + 1.3 \cdot |W_{tc}^R| \left|\dfrac{W_{tc}^L}{V_{cb}}\right| s_c\right)\left| \dfrac{A_t - \mu/\tan\beta}{m_{\tilde{t}}}\right| \left(\dfrac{1.5 \text{TeV}}{m_{\tilde{g}}} \right)^2,
 \label{eq:EDMn_SUSY}
\end{equation}
where $s_u$ and $s_c$ are the sines of the phases of $W_{tu}^R W_{tu}^L/V_{ub}$ and $W_{tc}^R W_{tc}^L/V_{cb}$ respectively. The sign ambiguity in the contribution of the Weinberg operator to $d_n$ can be reabsorbed in the sign of $s_c$.
Assuming the same matrix elements for the operators $\mathcal{O}_8$ and $\mathcal{O}_8'$ for simplicity, one can cast $\Delta$A$_{\rm CP}$ in the analogous form
\begin{equation}
\frac{\Delta\text{A}_{\rm CP}}{3.29 \cdot 10^{-3}} = \left(0.53 \cdot |W_{tc}^R| \left|\dfrac{W_{tu}^L}{V_{ub}}\right| s_8 + 6.17 \cdot |W_{tu}^R| \left|\dfrac{W_{tc}^L}{V_{cb}}\right| s_8^\prime\right) \left| \dfrac{A_t - \mu/\tan\beta}{m_{\tilde{t}}}\right| \left(\dfrac{1.5 \text{TeV}}{m_{\tilde{g}}} \right)^2,
 \label{eq:DaCP_SUSY}
\end{equation}
where $s_8$ and $s_8^\prime$ are the sines of the phases of $W_{tc}^R W_{tu}^L/V_{ub}$ and $W_{tu}^R W_{tc}^L/V_{cb}$ respectively. Also, the presence of a flavour blind phase in the mixing can easily be reabsorbed in the definitions of $s_{u,c}$, $s_8$ and $s_8^\prime$.
In Figure \ref{fig:SUSY_bounds} we show the bounds on $d_n$ and $\Delta$A$_{\rm CP}$ in the $|W_{tc}^R|$--$|W_{tu}^R|$ plane, for $m_{\tilde{g}} = 2 m_{\tilde{t}} = 1.5$ TeV and $(A_t - \mu/\tan \beta)/m_{\tilde{t}} = 1$. For illustrative purposes we assume all the phases to be maximal, and the left rotation $W^L$ elements to be equal in magnitude to the respective CKM ones. The generalization to the case where there are deviations from these reference values is easily readable off Eqs \eqref{eq:EDMn_SUSY} and \eqref{eq:DaCP_SUSY}. 

At present, for the reference values of the parameters in Eqs. \eqref{eq:SUSY_CEDMS_bounds} and \eqref{eq:SUSY_DeltaAcp_bounds}, the right charm-stop mixing angle $\theta_{ct}^R$ ($W_{tc}^R \simeq \cos\theta_{ct}^R \sin\theta_{ct}^R$) is not strongly constrained. In particular, values of $|W_{tc}^R| \gtrsim 0.3$ would both weaken the experimental lower bounds on the stop mass and mildly reduce fine-tuning \cite{Blanke:2013uia}. The projected sensitivities to EDMs shown in Table \ref{tab:exp_bounds} allow one to infer the impact of near-future experimental searches. If flavour violating phases are not significantly suppressed, a negative result at those experiments would reduce the allowed range for the charm-stop mixing by roughly two order of magnitudes\footnote{
A study of future collider sensitivity to $|W_{tc}^R|$ goes beyond the purposes of this work, see Ref. \cite{Blanke:2013uia} for an explorative analysis.}.

One could wonder when contributions from the exchange of squarks of the first two generations could interfere with the above ones, and affect in this way the bounds we derived. Those contributions are suppressed by a factor $y_{u,c}/y_t$ evaluated at the high scale, but are at the same time CKM enhanced if one normalize consistently the left mixing matrices $W^L$.
Because of this, and of the bound in \eqref{eq:SUSY_CEDMS_bounds}, the contribution to $\tilde{d}_u$ by a scharm circulating in the loop is potentially the larger one. We checked that, for the reference values for $m_{\tilde{g}}$ and $m_{\tilde{t}}$ that we chose, $m_{\tilde c} \gtrsim 5$~TeV is compatible with the bound on $\tilde{d}_u$ for Im$(W_{cu}^R)$ as large as 1. A smaller value of $m_{\tilde c}$ would imply a stronger bound on Im$(W_{cu}^R)$, and would in general affect the bound on $|W_{tu}^R|$ by $O(1)$. Thus it would affect the vertical axis of Fig \ref{fig:SUSY_bounds}, but it will not change the impact, on this picture, of the newly derived bound on $\tilde{d}_c$.
Its impact would of course be changed by a modification of the bound on $|W_{tc}^R|$. We checked that the same lower bound $m_{\tilde c} \gtrsim 5$~TeV implies that the contribution to $\tilde{d}_c$ is dominated by the third-generation diagram, until Im$(W_{tc}^R) \gtrsim 10^{-2} \text{Im}(W_{cc}^R)$. Thus, with these values of the masses, effects of the first two generation squarks would start to become relevant for the future reach of EDM experiments, if EDMs are still measured to be consistent with zero and if Im$(W_{cc}^R) \simeq 1$.

Finally notice that we have neglected the contribution that would come from CP violation in the Higgs and gaugino sectors, which in any case would also be constrained by the bound\cite{Baron:2013eja} on the electron EDM (see e.g. \cite{Barbieri:2014tja}).

\subsection{Composite Higgs models}
\label{sec:CHM}
It is interesting to see how the new bound on the charm CEDM impacts on composite Higgs models \cite{Kaplan:1983fs,Georgi:1984af,Contino:2003ve,Agashe:2004rs}, as a concrete realization of a dynamical suppression of flavour violating processes. We will in fact stick to partial compositeness \cite{Kaplan:1991dc} as a way to give masses to the SM quarks and to suppress at the same time flavour-changing neutral currents. We will consider a simplified two-site picture, in the spirit of \cite{Contino:2006nn}. In particular we will include one composite resonance for each SM boson and fermion field.
For the purpose of understanding the rest of this section, it is sufficient to define the following phenomenological Lagrangian for the strong sector\footnote{We understand canonically normalized kinetic terms
. Details about the parameters depend on the specific representation of the resonances, see \cite{Contino:2006nn} and, e.g., \cite{Barbieri:2012tu}.}
\begin{equation}
\mathcal{L}_{\rm s} = \frac{M_*^2}{2} \rho_\mu^2 - m_F \bar{F} F - (Y_U \bar{Q} H U + Y_D \bar{Q} H^c D + \text{h.c.}) - V(H)\,,
 \label{eq:L_CHM}
\end{equation}
and for the mixing of the composite fermions $F$ with the elementary ones $f$
\begin{equation}
\mathcal{L}_{\rm mix} = \lambda_L \bar{q}_LQ + \lambda_{Ru} \bar{U}u_R + \lambda_{Ru} \bar{D}d_R + \text{h.c.}\,.
 \label{eq:L_CHM_mix}
\end{equation}
Here $H$ denotes the Higgs doublet, and $\rho_\mu$ the composite vectors. Indices in flavour space are understood for the mixings $\lambda_f$, as well as for the composite Yukawas and fermion masses $Y_F$ and $m_F$. A sum over all species of fermionic and gauge fields is also understood. The mixings can always be brought to diagonal form, and rotated away in order to obtain the SM fields $f_{\rm SM} = \cos \theta_f f + \sin \theta_f F$
.

The dominant contributions to chirality breaking operators come from one-loop diagrams involving a fermion resonance and either the Higgs boson or the longitudinal component of $W$ and $Z$. In fact diagrams with a vector and a fermion resonances running in the loop have the same flavour and CP structure of the SM Yukawa terms. Thus they will be diagonal in flavour space, as well as real, in the mass basis for the SM fields. On the contrary the presence of two additional vertices with the composite Higgs introduces two extra composite Yukawas, which are anarchic in flavour space, giving rise to operators that are generically not aligned with the mass basis. Notice that a semiperturbative composite Yukawa coupling is preferred both by the Higgs mass value and naturalness arguments \cite{Contino:2006qr,Matsedonskyi:2012ym,Redi:2012ha,Pomarol:2012qf,Marzocca:2012zn}, as well as by precision constraints \cite{Barbieri:2012tu}, and thus a loop expansion in this coupling is not inconsistent.

The contributions to the Wilson coefficients of the up and charm CEDMs \eqref{EDM} and of $\Delta$A$_{\rm CP}$ \eqref{eq:Leff_Ddecay} are suppressed by
\begin{equation}
 \xi_{u,c} = \frac{y_{u,c}}{y_t} \, \big(Y^*_{u,c}\big)^2\,, \qquad \xi_8 = \lambda_C \, \frac{y_c}{y_t} \, \big(Y^*_8\big)^2\,, \qquad  \xi_8^{\prime} = \frac{y_u}{y_t}\frac{1}{\lambda_C} \, \big(Y^*_{8^\prime}\big)^2\,,
 \label{eq:CHM_supprCoeffs}
\end{equation}
where $\lambda_C$ is the Cabibbo angle, and $Y^*_{u,c}$, $Y^*_{8,8^\prime}$ are linear combinations of elements of the anarchic composite Yukawa matrices of \eqref{eq:L_CHM}, which are in general complex. Notice that those linear combinations depend also on which generations of composite resonances are running in the loop.

To simplify the discussion, it is convenient to decouple the first two generations of composite fermions. Naturalness considerations and the measured value of the Higgs mass require only the third generation resonances to lie close to the Fermi scale, while the other ones could well be heavier\footnote{Unlike in Supersymmetry, direct collider bounds are not an additional motivation to choose such a spectrum, see e.g. \cite{Redi:2013eaa}.} \cite{Contino:2006qr}.
This assumption implies the relation $Y^*_{8^\prime} Y^*_8 = Y^*_c Y^*_u$, and also that the order one coefficients of \eqref{EDM} and \eqref{eq:Leff_Ddecay} are all equal. They can be obtained from Refs. \cite{Agashe:2008uz,Vignaroli:2012si}, where the one-loop contribution coming from a fermion resonance running in the loop together with the Higgs and the Goldstone bosons is computed. Neglecting terms further suppressed by $O(m_W^2/m_T^2)$, we find
\begin{equation}
 \frac{c_D}{\Lambda^2} = \frac{c_D^\prime}{\Lambda^2} =\frac{\tilde c_c}{\Lambda^2} = \frac{\tilde c_u}{\Lambda^2} \simeq \frac{1}{16 \pi^2} \frac{9/8}{m_T^2}\,,
 \label{eq:CHM_O1Coeffs}
\end{equation}
where we assumed the partners of the left- and right-handed top and bottom quarks to have the same mass $m_T$. The bounds on the up and charm CEDMs then imply
\begin{equation}
  m_T \gtrsim 2.1 \,Y^*_u \;\text{TeV}, \qquad  m_T \gtrsim 1.2 \, Y^*_c \;\text{TeV}\,,\label{CHM:dn_T}
\end{equation}
to be compared with the ones coming from $\Delta$A$_{\rm CP}$
\begin{equation}
  m_T \gtrsim 1.3 \,Y^*_8 \;\text{TeV}, \qquad  m_T \gtrsim 0.26 \, Y^*_{8^\prime}\;\text{TeV}\,.\label{CHM:Dacp_T}
\end{equation}
Currently one could still saturate the experimental upper limit on $\Delta$A$_{\rm CP}$ in such scenarios \cite{KerenZur:2012fr,Delaunay:2012cz}, without running into any conflict with the bounds from the neutron EDM (for example by taking $Y_8^*$ sufficiently larger than $Y_u^*$). With the foreseen improvement in experimental sensitivity this possibility will be strongly challenged, for semiperturbative values of the composite Yukawas.
Notice also that the combination $Y^*_{8^\prime}  Y^*_8 = Y^*_c Y^*_u$ is more constrained by the bounds from the CEDMs than from those coming from $\Delta$A$_{\rm CP}$.

An analysis of the total contribution to $d_n$, similar to the one performed in Section \ref{sec:SUSY} for Supersymmetry, cannot be carried out in an analogously simple way for CHMs. This is due mainly to the presence of potentially unsuppressed contributions to $d_n$ from $d_d$ and $\tilde{d}_d$.

\section{Summary and conclusions}
\label{sec:Summary}
Measurements of CP violating observables are among the strongest indirect probes of high energy scales. It is therefore important to study their implications for our knowledge of physics beyond the SM. In this paper, we pursued a step in the above direction.

We derived bounds on the charm electric and chromo-electric dipole moments, $d_c$ and $\tilde{d}_c$. For $d_c$, we considered its possibile dangerous contributions to the neutron EDM, $d_n$, and to the branching ratio BR($B \to X_s \gamma$).
In the first case we made use of the contribution of $d_c$ to $d_d$ from electroweak running, and derived the bound in Eq. \eqref{eq:EDMc_bound_dn}. In the second case we considered the relevant loop process proportional to $d_c$, yielding to the bound in Eq. \eqref{eq:EDMc_bound_Bsgamma}. However, the stronger bound was by far the one on the charm CEDM $\tilde{d}_c$. We obtained it via its threshold effect in the three gluon Weinberg operator. This operator in turn contributes to hadronic dipole moments, like the neutron and the deuteron ones, yielding to $\tilde{d}_c < 1.0 \times 10^{-22}\,e$ cm  at 90\% C.L. at the charm mass scale. This is one of the two main results of this paper.

We also pointed out the relevance of this bound for models allowing for a non-negligible flavour violation in the right-handed up quarks sector. These models are still largely unconstrained due to the weakness of the flavour and CP violating bounds compared to those for the down-quark sector. Explicit examples are models of flavour alignment and Generic $U(2)^3$. 
Before this work, the CP asymmetry in flavour violating $D$ decays, $\Delta$A$_{\rm CP}$, was setting the stronger constraints on the relevant flavour violating parameters in these models. We found that the current bound on $d_n$ is already sligthly more constraining than $\Delta$A$_{\rm CP}$. More importantly, the lack of a theoretical understanding of the SM contribution to $\Delta$A$_{\rm CP}$, combined with the expected improvement in experimental sensitivity to $d_n$, will make the neutron EDM the most sensitive probe for these flavour violating parameters, strengthening the current bounds by more than two orders of magnitude.
We also specialized our analysis to various new physics models, such as split-families Supersymmetry, and composite Higgs models with partial compositeness. In particular in the first case, under some motivated assumptions, it was possible to find concise expressions for the total supersymmetric contribution to both $d_n$ and $\Delta$A$_{\rm CP}$.

We think that these results constitute a further motivation to increase the experimental sensitivities to $d_n$ and $\Delta$A$_{\rm CP}$, and to continue the effort to achieve a better theoretical control of the latter.

\subsubsection*{Acknowledgments}
We thank Michele Papucci for many precious discussions and for comments on the manuscript. We also thank Jordy de Vries and Emanuele Mereghetti for useful discussions, and Martin Gorbahn and Ulrich Haisch for spotting an error (now corrected) in Eq. \eqref{eq:EDMc_bound_Bsgamma}.
This work is 
supported in part by the European Programme ``Unification in the LHC Era", contract PITN-GA-2009-237920 (UNI\-LHC), by MIUR under the contract 2010YJ2NYW-010, and by the European Research Council (ERC) under the EU Seventh Framework Programme (FP7/2007-2013) / {\sc Erc} Starting Grant (agreement n. 278234 - ‘{\sc NewDark}’ project).

\bibliographystyle{My}
\small
\bibliography{Dipoles}

\providecommand{\href}[2]{#2}\begingroup\raggedright\begin{thebibliography}{10}

\bibitem{Hewett:2012ns}
J.~Hewett, H.~Weerts, R.~Brock, J.~Butler, B.~Casey, {\em et al.,}
\href{http://arxiv.org/abs/1205.2671}{{\tt arXiv:1205.2671 [hep-ex]}}.

\bibitem{Baker:2006ts}
C.~Baker, D.~Doyle, P.~Geltenbort, K.~Green, M.~van~der Grinten, {\em et al.,}
  \href{http://dx.doi.org/10.1103/PhysRevLett.97.131801}{{\em Phys.Rev.Lett.}
  {\bf 97} (2006)  131801},
\href{http://arxiv.org/abs/hep-ex/0602020}{{\tt arXiv:hep-ex/0602020
  [hep-ex]}}.

\bibitem{Griffith:2009zz}
W.~Griffith, M.~Swallows, T.~Loftus, M.~Romalis, B.~Heckel, {\em et al.,}
\href{http://dx.doi.org/10.1103/PhysRevLett.102.101601}{{\em Phys.Rev.Lett.}
  {\bf 102} (2009)  101601}.

\bibitem{Bodek:2008gr}
K.~Bodek, S.~Kistryn, M.~Kuzniak, J.~Zejma, M.~Burghoff, {\em et al.,}
\href{http://arxiv.org/abs/0806.4837}{{\tt arXiv:0806.4837 [nucl-ex]}}.

\bibitem{Altarev:2009zz}
I.~Altarev, G.~Ban, G.~Bison, K.~Bodek, M.~Burghoff, {\em et al.,}
\href{http://dx.doi.org/10.1016/j.nima.2009.07.046}{{\em Nucl.Instrum.Meth.}
  {\bf A611} (2009)  133--136}.

\bibitem{Baker:2010zza}
C.~Baker, S.~Balashov, V.~Francis, K.~Green, M.~van~der Grinten, {\em et al.,}
\href{http://dx.doi.org/10.1088/1742-6596/251/1/012055}{{\em J.Phys.Conf.Ser.}
  {\bf 251} (2010)  012055}.

\bibitem{Beck:2011gw}
{\bf nEDM Collaboration}, D.~Beck, D.~Budker, and B.~Park,
\href{http://arxiv.org/abs/1111.1273}{{\tt arXiv:1111.1273 [nucl-ex]}}.

\bibitem{Altarev:2012uy}
I.~Altarev, D.~Beck, S.~Chesnevskaya, T.~Chupp, W.~Feldmeier, {\em et al.,}
\href{http://dx.doi.org/10.1393/ncc/i2012-11271-0}{{\em Nuovo Cim.} {\bf
  C035N04} (2012)  122--127}.

\bibitem{bnl:gov}
http://www.bnl.gov/edm/.

\bibitem{Shabalin:1978rs}
E.~Shabalin
{\em Sov.J.Nucl.Phys.} {\bf 28} (1978)  75.

\bibitem{Khriplovich:1985jr}
I.~Khriplovich
\href{http://dx.doi.org/10.1016/0370-2693(86)90245-5}{{\em Phys.Lett.} {\bf
  B173} (1986)  193--196}.

\bibitem{Czarnecki:1997bu}
A.~Czarnecki and B.~Krause,
  \href{http://dx.doi.org/10.1103/PhysRevLett.78.4339}{{\em Phys.Rev.Lett.}
  {\bf 78} (1997)  4339--4342},
\href{http://arxiv.org/abs/hep-ph/9704355}{{\tt arXiv:hep-ph/9704355
  [hep-ph]}}.

\bibitem{Mannel:2012qk}
T.~Mannel and N.~Uraltsev,
  \href{http://dx.doi.org/10.1103/PhysRevD.85.096002}{{\em Phys.Rev.} {\bf D85}
  (2012)  096002},
\href{http://arxiv.org/abs/1202.6270}{{\tt arXiv:1202.6270 [hep-ph]}}.

\bibitem{Pospelov:2000bw}
M.~Pospelov and A.~Ritz,
  \href{http://dx.doi.org/10.1103/PhysRevD.63.073015}{{\em Phys.Rev.} {\bf D63}
  (2001)  073015},
\href{http://arxiv.org/abs/hep-ph/0010037}{{\tt arXiv:hep-ph/0010037
  [hep-ph]}}.

\bibitem{Lebedev:2004va}
O.~Lebedev, K.~A. Olive, M.~Pospelov, and A.~Ritz,
  \href{http://dx.doi.org/10.1103/PhysRevD.70.016003}{{\em Phys.Rev.} {\bf D70}
  (2004)  016003},
\href{http://arxiv.org/abs/hep-ph/0402023}{{\tt arXiv:hep-ph/0402023
  [hep-ph]}}.

\bibitem{deVries:2011re}
J.~de~Vries, E.~Mereghetti, R.~Timmermans, and U.~van Kolck,
  \href{http://dx.doi.org/10.1103/PhysRevLett.107.091804}{{\em Phys.Rev.Lett.}
  {\bf 107} (2011)  091804},
\href{http://arxiv.org/abs/1102.4068}{{\tt arXiv:1102.4068 [hep-ph]}}.

\bibitem{deVries:2011an}
J.~de~Vries, R.~Higa, C.-P. Liu, E.~Mereghetti, I.~Stetcu, {\em et al.,}
  \href{http://dx.doi.org/10.1103/PhysRevC.84.065501}{{\em Phys.Rev.} {\bf C84}
  (2011)  065501},
\href{http://arxiv.org/abs/1109.3604}{{\tt arXiv:1109.3604 [hep-ph]}}.

\bibitem{Ellis:2008zy}
J.~R. Ellis, J.~S. Lee, and A.~Pilaftsis,
  \href{http://dx.doi.org/10.1088/1126-6708/2008/10/049}{{\em JHEP} {\bf 0810}
  (2008)  049},
\href{http://arxiv.org/abs/0808.1819}{{\tt arXiv:0808.1819 [hep-ph]}}.

\bibitem{Fuyuto:2012yf}
K.~Fuyuto, J.~Hisano, and N.~Nagata,
  \href{http://dx.doi.org/10.1103/PhysRevD.87.054018}{{\em Phys.Rev.} {\bf D87}
  (2013) no.~5, 054018},
\href{http://arxiv.org/abs/1211.5228}{{\tt arXiv:1211.5228 [hep-ph]}}.

\bibitem{Jung:2013hka}
M.~Jung and A.~Pich,
\href{http://arxiv.org/abs/1308.6283}{{\tt arXiv:1308.6283 [hep-ph]}}.

\bibitem{Chang:1990jv}
D.~Chang, W.-Y. Keung, C.~Li, and T.~Yuan,
\href{http://dx.doi.org/10.1016/0370-2693(90)91875-C}{{\em Phys.Lett.} {\bf
  B241} (1990)  589}.

\bibitem{Boyd:1990bx}
G.~Boyd, A.~K. Gupta, S.~P. Trivedi, and M.~B. Wise,
\href{http://dx.doi.org/10.1016/0370-2693(90)91874-B}{{\em Phys.Lett.} {\bf
  B241} (1990)  584}.

\bibitem{Dine:1990pf}
M.~Dine and W.~Fischler,
\href{http://dx.doi.org/10.1016/0370-2693(90)91464-M}{{\em Phys.Lett.} {\bf
  B242} (1990)  239--244}.

\bibitem{Braaten:1990gq}
E.~Braaten, C.-S. Li, and T.-C. Yuan,
\href{http://dx.doi.org/10.1103/PhysRevLett.64.1709}{{\em Phys.Rev.Lett.} {\bf
  64} (1990)  1709}.

\bibitem{Degrassi:2005zd}
G.~Degrassi, E.~Franco, S.~Marchetti, and L.~Silvestrini,
  \href{http://dx.doi.org/10.1088/1126-6708/2005/11/044}{{\em JHEP} {\bf 0511}
  (2005)  044},
\href{http://arxiv.org/abs/hep-ph/0510137}{{\tt arXiv:hep-ph/0510137
  [hep-ph]}}.

\bibitem{Kuang:2012wp}
Y.-P. Kuang, J.-P. Ma, O.~Nachtmann, W.-P. Xie, and H.-H. Zheng,
  \href{http://dx.doi.org/10.1103/PhysRevD.85.114010}{{\em Phys.Rev.} {\bf D85}
  (2012)  114010},
\href{http://arxiv.org/abs/1202.3042}{{\tt arXiv:1202.3042 [hep-ph]}}.

\bibitem{Kamenik:2011dk}
J.~F. Kamenik, M.~Papucci, and A.~Weiler,
  \href{http://dx.doi.org/10.1103/PhysRevD.85.071501}{{\em Phys.Rev.} {\bf D85}
  (2012)  071501},
\href{http://arxiv.org/abs/1107.3143}{{\tt arXiv:1107.3143 [hep-ph]}}.

\bibitem{CorderoCid:2007uc}
A.~Cordero-Cid, J.~Hernandez, G.~Tavares-Velasco, and J.~Toscano,
  \href{http://dx.doi.org/10.1088/0954-3899/35/2/025004}{{\em J.Phys.} {\bf
  G35} (2008)  025004},
\href{http://arxiv.org/abs/0712.0154}{{\tt arXiv:0712.0154 [hep-ph]}}.

\bibitem{Hewett:1993em}
J.~L. Hewett and T.~G. Rizzo,
  \href{http://dx.doi.org/10.1103/PhysRevD.49.319}{{\em Phys.Rev.} {\bf D49}
  (1994)  319--322},
\href{http://arxiv.org/abs/hep-ph/9305223}{{\tt arXiv:hep-ph/9305223
  [hep-ph]}}.

\bibitem{Buras:2011zb}
A.~J. Buras, L.~Merlo, and E.~Stamou,
  \href{http://dx.doi.org/10.1007/JHEP08(2011)124}{{\em JHEP} {\bf 1108} (2011)
   124},
\href{http://arxiv.org/abs/1105.5146}{{\tt arXiv:1105.5146 [hep-ph]}}.

\bibitem{Amhis:2012bh}
{\bf Heavy Flavor Averaging Group}, Y.~Amhis {\em et al.,}
  \href{http://arxiv.org/abs/1207.1158}{{\tt arXiv:1207.1158 [hep-ex]}}.
Online update at http://www.slac.stanford.edu/xorg/hfag.

\bibitem{Misiak:2006zs}
M.~Misiak, H.~Asatrian, K.~Bieri, M.~Czakon, A.~Czarnecki, {\em et al.,}
  \href{http://dx.doi.org/10.1103/PhysRevLett.98.022002}{{\em Phys.Rev.Lett.}
  {\bf 98} (2007)  022002},
\href{http://arxiv.org/abs/hep-ph/0609232}{{\tt arXiv:hep-ph/0609232
  [hep-ph]}}.

\bibitem{Hisano:2012cc}
J.~Hisano, K.~Tsumura, and M.~J. Yang,
  \href{http://dx.doi.org/10.1016/j.physletb.2012.06.038}{{\em Phys.Lett.} {\bf
  B713} (2012)  473--480},
\href{http://arxiv.org/abs/1205.2212}{{\tt arXiv:1205.2212 [hep-ph]}}.

\bibitem{Nir:1993mx}
Y.~Nir and N.~Seiberg,
  \href{http://dx.doi.org/10.1016/0370-2693(93)90942-B}{{\em Phys.Lett.} {\bf
  B309} (1993)  337--343},
\href{http://arxiv.org/abs/hep-ph/9304307}{{\tt arXiv:hep-ph/9304307
  [hep-ph]}}.

\bibitem{Barbieri:2012bh}
R.~Barbieri, D.~Buttazzo, F.~Sala, and D.~M. Straub,
  \href{http://dx.doi.org/10.1007/JHEP10(2012)040}{{\em JHEP} {\bf 1210} (2012)
   040},
\href{http://arxiv.org/abs/1206.1327}{{\tt arXiv:1206.1327 [hep-ph]}}.

\bibitem{Isidori:2011qw}
G.~Isidori, J.~F. Kamenik, Z.~Ligeti, and G.~Perez,
  \href{http://dx.doi.org/10.1016/j.physletb.2012.03.046}{{\em Phys.Lett.} {\bf
  B711} (2012)  46--51},
\href{http://arxiv.org/abs/1111.4987}{{\tt arXiv:1111.4987 [hep-ph]}}.

\bibitem{Aaij:2013bra}
{\bf LHCb collaboration}, R.~Aaij {\em et al.,}
  \href{http://dx.doi.org/10.1016/j.physletb.2013.04.061}{{\em Phys.Lett.} {\bf
  B723} (2013)  33--43},
\href{http://arxiv.org/abs/1303.2614}{{\tt arXiv:1303.2614 [hep-ex]}}.

\bibitem{Pirtskhalava:2011va}
D.~Pirtskhalava and P.~Uttayarat,
  \href{http://dx.doi.org/10.1016/j.physletb.2012.04.039}{{\em Phys.Lett.} {\bf
  B712} (2012)  81--86},
\href{http://arxiv.org/abs/1112.5451}{{\tt arXiv:1112.5451 [hep-ph]}}.

\bibitem{Cheng:2012wr}
H.-Y. Cheng and C.-W. Chiang,
  \href{http://dx.doi.org/10.1103/PhysRevD.85.079903,
  10.1103/PhysRevD.85.034036}{{\em Phys.Rev.} {\bf D85} (2012)  034036},
\href{http://arxiv.org/abs/1201.0785}{{\tt arXiv:1201.0785 [hep-ph]}}.

\bibitem{Brod:2012ud}
J.~Brod, Y.~Grossman, A.~L. Kagan, and J.~Zupan,
\href{http://arxiv.org/abs/1203.6659}{{\tt arXiv:1203.6659 [hep-ph]}}.

\bibitem{Isidori:2012yx}
G.~Isidori and J.~F. Kamenik,
\href{http://arxiv.org/abs/1205.3164}{{\tt arXiv:1205.3164 [hep-ph]}}.

\bibitem{Giudice:2012qq}
G.~F. Giudice, G.~Isidori, and P.~Paradisi,
  \href{http://dx.doi.org/10.1007/JHEP04(2012)060}{{\em JHEP} {\bf 1204} (2012)
   060},
\href{http://arxiv.org/abs/1201.6204}{{\tt arXiv:1201.6204 [hep-ph]}}.

\bibitem{Mannel:2012hb}
T.~Mannel and N.~Uraltsev,
  \href{http://dx.doi.org/10.1007/JHEP03(2013)064}{{\em JHEP} {\bf 1303} (2013)
   064},
\href{http://arxiv.org/abs/1205.0233}{{\tt arXiv:1205.0233 [hep-ph]}}.

\bibitem{Dimopoulos:1995mi}
S.~Dimopoulos and G.~Giudice,
  \href{http://dx.doi.org/10.1016/0370-2693(95)00961-J}{{\em Phys.Lett.} {\bf
  B357} (1995)  573--578},
\href{http://arxiv.org/abs/hep-ph/9507282}{{\tt arXiv:hep-ph/9507282
  [hep-ph]}}.

\bibitem{Cohen:1996vb}
A.~G. Cohen, D.~Kaplan, and A.~Nelson,
  \href{http://dx.doi.org/10.1016/S0370-2693(96)01183-5}{{\em Phys.Lett.} {\bf
  B388} (1996)  588--598},
\href{http://arxiv.org/abs/hep-ph/9607394}{{\tt arXiv:hep-ph/9607394
  [hep-ph]}}.

\bibitem{Barbieri:2009ev}
R.~Barbieri and D.~Pappadopulo,
  \href{http://dx.doi.org/10.1088/1126-6708/2009/10/061}{{\em JHEP} {\bf 0910}
  (2009)  061},
\href{http://arxiv.org/abs/0906.4546}{{\tt arXiv:0906.4546 [hep-ph]}}.

\bibitem{Papucci:2011wy}
M.~Papucci, J.~T. Ruderman, and A.~Weiler,
  \href{http://dx.doi.org/10.1007/JHEP09(2012)035}{{\em JHEP} {\bf 1209} (2012)
   035},
\href{http://arxiv.org/abs/1110.6926}{{\tt arXiv:1110.6926 [hep-ph]}}.

\bibitem{Blanke:2013uia}
M.~Blanke, G.~F. Giudice, P.~Paradisi, G.~Perez, and J.~Zupan,
  \href{http://dx.doi.org/10.1007/JHEP06(2013)022}{{\em JHEP} {\bf 1306} (2013)
   022},
\href{http://arxiv.org/abs/1302.7232}{{\tt arXiv:1302.7232 [hep-ph]}}.

\bibitem{Baron:2013eja}
{\bf ACME Collaboration}, J.~Baron {\em et al.,}
\href{http://arxiv.org/abs/1310.7534}{{\tt arXiv:1310.7534 [physics.atom-ph]}}.

\bibitem{Barbieri:2014tja}
R.~Barbieri, D.~Buttazzo, F.~Sala, and D.~M. Straub,
\href{http://arxiv.org/abs/1402.6677}{{\tt arXiv:1402.6677 [hep-ph]}}.

\bibitem{Kaplan:1983fs}
D.~B. Kaplan and H.~Georgi,
\href{http://dx.doi.org/10.1016/0370-2693(84)91177-8}{{\em Phys.Lett.} {\bf
  B136} (1984)  183}.

\bibitem{Georgi:1984af}
H.~Georgi and D.~B. Kaplan,
\href{http://dx.doi.org/10.1016/0370-2693(84)90341-1}{{\em Phys.Lett.} {\bf
  B145} (1984)  216}.

\bibitem{Contino:2003ve}
R.~Contino, Y.~Nomura, and A.~Pomarol,
  \href{http://dx.doi.org/10.1016/j.nuclphysb.2003.08.027}{{\em Nucl.Phys.}
  {\bf B671} (2003)  148--174},
\href{http://arxiv.org/abs/hep-ph/0306259}{{\tt arXiv:hep-ph/0306259
  [hep-ph]}}.

\bibitem{Agashe:2004rs}
K.~Agashe, R.~Contino, and A.~Pomarol,
  \href{http://dx.doi.org/10.1016/j.nuclphysb.2005.04.035}{{\em Nucl.Phys.}
  {\bf B719} (2005)  165--187},
\href{http://arxiv.org/abs/hep-ph/0412089}{{\tt arXiv:hep-ph/0412089
  [hep-ph]}}.

\bibitem{Kaplan:1991dc}
D.~B. Kaplan \href{http://dx.doi.org/10.1016/S0550-3213(05)80021-5}{{\em
  Nucl.Phys.} {\bf B365} (1991)  259--278}.
Revised version.

\bibitem{Contino:2006nn}
R.~Contino, T.~Kramer, M.~Son, and R.~Sundrum,
  \href{http://dx.doi.org/10.1088/1126-6708/2007/05/074}{{\em JHEP} {\bf 0705}
  (2007)  074},
\href{http://arxiv.org/abs/hep-ph/0612180}{{\tt arXiv:hep-ph/0612180
  [hep-ph]}}.

\bibitem{Barbieri:2012tu}
R.~Barbieri, D.~Buttazzo, F.~Sala, D.~M. Straub, and A.~Tesi,
  \href{http://dx.doi.org/10.1007/JHEP05(2013)069}{{\em JHEP} {\bf 1305} (2013)
   069},
\href{http://arxiv.org/abs/1211.5085}{{\tt arXiv:1211.5085 [hep-ph]}}.

\bibitem{Contino:2006qr}
R.~Contino, L.~Da~Rold, and A.~Pomarol,
  \href{http://dx.doi.org/10.1103/PhysRevD.75.055014}{{\em Phys.Rev.} {\bf D75}
  (2007)  055014},
\href{http://arxiv.org/abs/hep-ph/0612048}{{\tt arXiv:hep-ph/0612048
  [hep-ph]}}.

\bibitem{Matsedonskyi:2012ym}
O.~Matsedonskyi, G.~Panico, and A.~Wulzer,
  \href{http://dx.doi.org/10.1007/JHEP01(2013)164}{{\em JHEP} {\bf 1301} (2013)
   164},
\href{http://arxiv.org/abs/1204.6333}{{\tt arXiv:1204.6333 [hep-ph]}}.

\bibitem{Redi:2012ha}
M.~Redi and A.~Tesi, \href{http://dx.doi.org/10.1007/JHEP10(2012)166}{{\em
  JHEP} {\bf 1210} (2012)  166},
\href{http://arxiv.org/abs/1205.0232}{{\tt arXiv:1205.0232 [hep-ph]}}.

\bibitem{Pomarol:2012qf}
A.~Pomarol and F.~Riva, \href{http://dx.doi.org/10.1007/JHEP08(2012)135}{{\em
  JHEP} {\bf 1208} (2012)  135},
\href{http://arxiv.org/abs/1205.6434}{{\tt arXiv:1205.6434 [hep-ph]}}.

\bibitem{Marzocca:2012zn}
D.~Marzocca, M.~Serone, and J.~Shu,
  \href{http://dx.doi.org/10.1007/JHEP08(2012)013}{{\em JHEP} {\bf 1208} (2012)
   013},
\href{http://arxiv.org/abs/1205.0770}{{\tt arXiv:1205.0770 [hep-ph]}}.

\bibitem{Redi:2013eaa}
M.~Redi, V.~Sanz, M.~de~Vries, and A.~Weiler,
  \href{http://dx.doi.org/10.1007/JHEP08(2013)008}{{\em JHEP} {\bf 1308} (2013)
   008},
\href{http://arxiv.org/abs/1305.3818}{{\tt arXiv:1305.3818 [hep-ph]}}.

\bibitem{Agashe:2008uz}
K.~Agashe, A.~Azatov, and L.~Zhu,
  \href{http://dx.doi.org/10.1103/PhysRevD.79.056006}{{\em Phys.Rev.} {\bf D79}
  (2009)  056006},
\href{http://arxiv.org/abs/0810.1016}{{\tt arXiv:0810.1016 [hep-ph]}}.

\bibitem{Vignaroli:2012si}
N.~Vignaroli \href{http://dx.doi.org/10.1103/PhysRevD.86.115011}{{\em
  Phys.Rev.} {\bf D86} (2012)  115011},
\href{http://arxiv.org/abs/1204.0478}{{\tt arXiv:1204.0478 [hep-ph]}}.

\bibitem{KerenZur:2012fr}
B.~Keren-Zur, P.~Lodone, M.~Nardecchia, D.~Pappadopulo, R.~Rattazzi, {\em et
  al.,} \href{http://dx.doi.org/10.1016/j.nuclphysb.2012.10.012}{{\em
  Nucl.Phys.} {\bf B867} (2013)  429--447},
\href{http://arxiv.org/abs/1205.5803}{{\tt arXiv:1205.5803 [hep-ph]}}.

\bibitem{Delaunay:2012cz}
C.~Delaunay, J.~F. Kamenik, G.~Perez, and L.~Randall,
  \href{http://dx.doi.org/10.1007/JHEP01(2013)027}{{\em JHEP} {\bf 1301} (2013)
   027},
\href{http://arxiv.org/abs/1207.0474}{{\tt arXiv:1207.0474 [hep-ph]}}.

\end{thebibliography}\endgroup

%
\end{document}